\documentclass[12pt]{article}
\usepackage{lineno}
\usepackage{graphicx}
\usepackage{amsmath}
\usepackage{latexsym}
\usepackage{amssymb}
\usepackage{gensymb}
\usepackage{nicefrac}

\newcommand\pubnumber{}
\newcommand\pubdate{\today}

\def\napoli{Department of Physics\\
Duke University, Durham, North Carolina 27708}

\def\Title#1{\begin{center} {\Large #1 } \end{center}}
\def\Author#1{\begin{center}{ \sc #1} \end{center}}
\def\Address#1{\begin{center}{ \it #1} \end{center}}

\newcommand\pubblock{\rightline{\begin{tabular}{l} \pubnumber\\
         \pubdate  \end{tabular}}}
\newenvironment{Abstract}{\begin{quotation}  }{\end{quotation}}
\newenvironment{Presented}{\begin{quotation} \begin{center} 
             PRESENTED AT\end{center}\bigskip 
      \begin{center}\begin{large}}{\end{large}\end{center} \end{quotation}}

\def\beq{\begin{equation}}
\def\eeq#1{\label{#1}\end{equation}}
\def\eeqn{\end{equation}}

\def\beqa{\begin{eqnarray}}
\def\eeqa#1{\label{#1}\end{eqnarray}}
\def\eeqan{\end{eqnarray}}

\let\bar=\overbar

\def\Dslash{\not{\hbox{\kern-4pt $D$}}}
\def\dslash{\not{\hbox{\kern-2pt $\del$}}}

\def\msb{{\bar{\ssstyle M \kern -1pt S}}}

\begin{document}
\begin{titlepage}
\pubblock

\vfill
\Title{Novel experimental probes of QCD in SIDIS and $e^+e^-$ annihilation}
\vfill
\Author{Anselm Vossen\\for the CLAS collaboration}
\Address{\napoli}
\vfill
\begin{Abstract}
Semi-inclusive deep inelastic Scattering (SIDIS) has been a very
successful tool to investigate the partonic structure of the nucleon
over the last decade.
Compared to inclusive DIS, information about the quantum numbers of
the struck quark can be inferred from the identity, distribution and
polarization of the final state hadrons.
Up to now, virtually all knowledge about the quark-gluon structure of
the nucleon from SIDIS has been gained from distributions of
independently observed
scalar hadrons.
However, given the amount of data current and future experiments at
JLab, RHIC, KEK and the EIC will collect, new paradigms have to be
explored to leverage the statistical power of the data. Similar to
other felds in nuclear and particle physics, it is natural to move
towards the exploration of more complex correlations in the observed
fi nal state.
This contribution will discuss recent results and future prospects of using
di-hadron correlations and polarized hyperon probes to study QCD in
SIDIS, pp and e+e- annihilation. Both of these probes exploit
additional degrees of freedom in the final state, given by the
relative momentum of the di-hadron pair and the hyperon polarization,
respectively.
This contribution will focus on recent results and opportunities opened by
these probes to study nucleon structure, hadronization and QCD in
novel ways. The focus will be on planned SIDIS measurements at CLAS12
at JLab and e+e- at Belle II.

\end{Abstract}
\vfill
\begin{Presented}
Thirteenth Conference on the Intersections of Particle and Nuclear Physics (CIPANP 2018)
\end{Presented}
\vfill
\end{titlepage}
\def\thefootnote{\fnsymbol{footnote}}
\setcounter{footnote}{0}

\section{Introduction}
In the process of Semi-Inclusive Deep Inelastic Scattering (SIDIS), a leptonic probe, usually a electron or muon, scatters off a target, {\textit e.g.} a proton, and the outgoing lepton with part of the hadronic final state are detected. SIDIS experiments have been extraordinarily successful in the exploration of the partonic structure of the nucleon, see {\textit e.g.}~Ref.~\cite{Aidala:2012mv}.
Up to now, the emphasis has been on SIDIS experiments where a single, scalar hadron, usually a pion or kaon, has been detected in the final state together with the scattered lepton.
It is quite obvious, that more can be learned about the nucleon structure and QCD dynamics in general, if final states are considered that can carry more information on the struck quark. Here, two such probes, di-hadron correlations and polarized $\Lambda$ production, are presented in some more detail along with a physics program that these probes enable at CLAS12 and Belle II. Both are new detectors, which started taking data this year. Due to their large acceptance, high precision tracking and particle identification capabilities, they are ideally suited for the di-hadron correlation and $\Lambda$ hyperon studies discussed here.
This contribution is structured as follows: Section~\ref{sec:diHadronCorrelations} introduces di-hadron correlations, with Sec.~\ref{sec:H1Perp} discussing the polarization dependent fragmentation function and Sec~\ref{sec:twist3} the twist-3 PDF that can be accessed in asymmetry measurements.
Section~\ref{sec:lambda} gives a short introduction why looking at polarized hyperons is interesting. Then, Sec.~\ref{sec:clas12} introduces the Clas12 experiment, with Sec.~\ref{sec:plannedClas12Measurements} outlining some planned measurements. The Belle II physics program is outlined very briefly in Sec.~\ref{sec:PlannedBelle}.

\section{Di-hadron Correlations}
\label{sec:diHadronCorrelations}
\begin{figure}
\begin{center}
\includegraphics[height=0.2\textheight,width=0.49\textwidth]{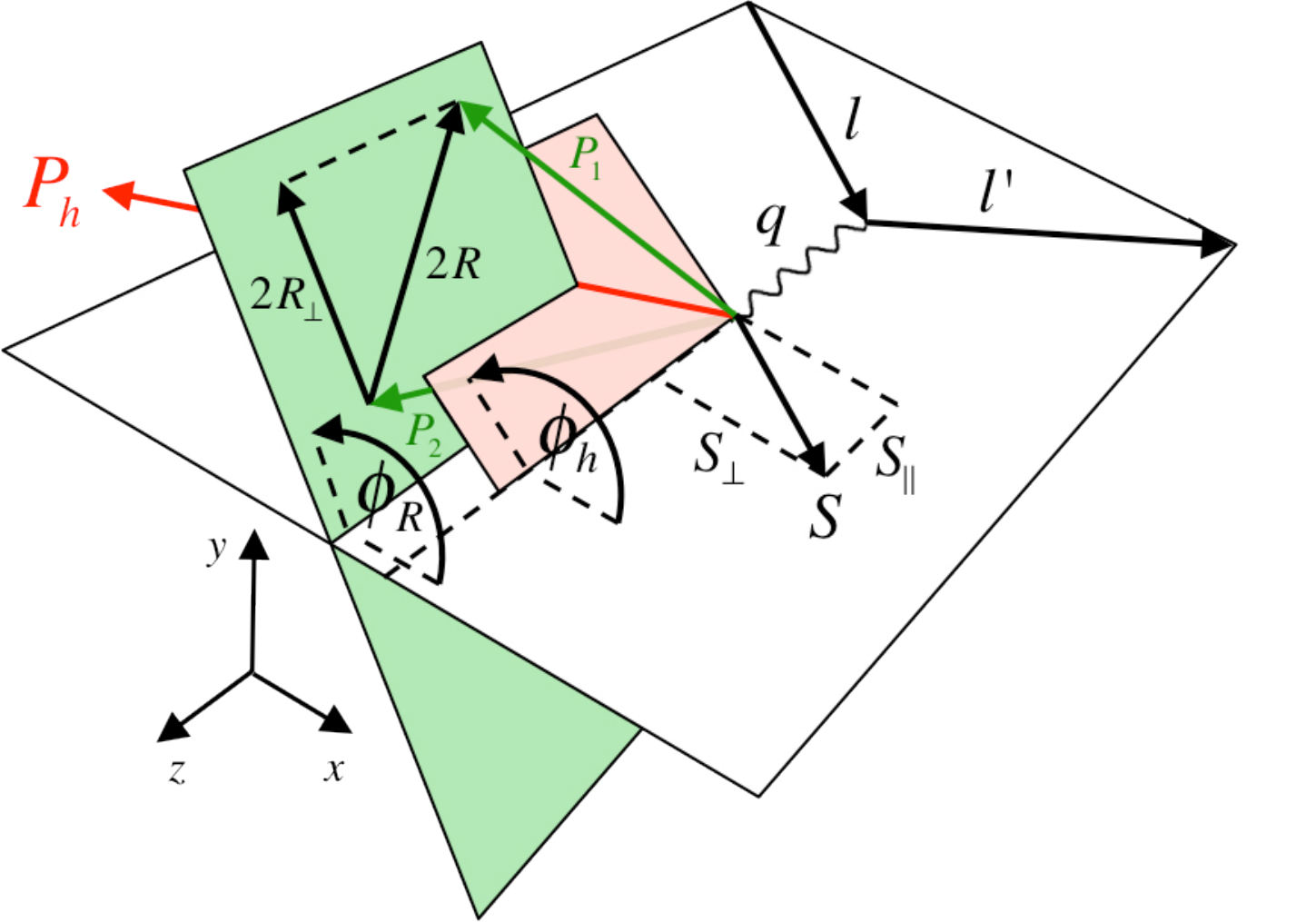}
\end{center}
\caption{Relevant angles and vectors for the measurement of di-hadron correlations. See text for more details. Figure taken from~\cite{Sirtl:2017rhi}\label{fig:2hframe}}
\end{figure}

The process $e(l)+p(P)\rightarrow e(l')+h_1(P_1)+h_2(P_2)+X$ is considered, where an electron with four-momentum $l$ is scattered off a target with four-momentum $P$ and two hadrons ($h_1,h_2$) with four-momenta $P_1$ and $P_2$ as well as the scattered electron are detected in the final state. 
The relevant kinematic variables for this process are the usual SIDIS variables $Q^2=-q^\mu q_\mu$, $\nu=P^\mu q_\mu/M_N$, $x=Q^2/(2M_N\nu)$ and $z=P_\mu P_h^\mu/(P^\mu q_\mu)$, where $M_N$ is the target mass and $P_h=P_1+P_2$ is the four-momentum of the hadron pair and $q$ the four-vector of the virtual photon momentum. The additional degree of freedom due to the second hadron is usually parametrized as $M_h$, the invariant mass of the two-hadron system, and the three-vector $\vec{R}=\vec{P_1}-\vec{P_2}$. 
Using QCD factorization theorems, the SIDIS cross-section can be written as a convolution of parton distribution functions (PDFs) and fragmentation functions (FFs). While the PDF describes the dynamics of quark and gluon fields inside the nucleon, the FFs, at leading twist, describe the probability of forming a hadron with specific quantum numbers from a given quark.  For a review on FFs, see Ref.~\cite{Metz:2016swz}.
For large $M_h$ the two-hadron fragmentation process can be derived perturbatively from single-hadron FFs, whereas for $M_h$ below the hard scale (at CLAS12 $\approx 1 GeV/c^2$) genuinely new non-perturbative objects are needed to describe the cross section, the so-called {\textit Di-hadron Fragmentation Functions (DiFFs)}. The  dependence of fragmentation function on the polarization of the initial quarks leads to a dependence of the FF on the azimuthal angles
 of $\vec{P}_h$ and $\vec{R}$ in the Breit frame, $\phi_h$ and $\phi_R$. These angles are depicted in Fig.~\ref{fig:2hframe}.
\subsection{Collinear di-hadron FF $H_1^\sphericalangle$}
\label{sec:H1Perp}
When integrated over $\phi_h$, DiFFs are collinear objects but, unlike single hadron FFs, they can still carry angular momentum and thus enable measurements sensitive to the transverse polarization of partons. 
The respective chiral-odd DiFF is for historical reasons often referred to as "Interference-Fragmentation Function" (IFF) and is denoted as $H_1^\sphericalangle$. Due to its importance in studies of the transverse spin structure of the nucleon, it has been the focus of substantial theoretical and experimental work. It was first measured at Belle~\cite{Vossen:2011fk}, which enabled the first extraction of $H_1^\sphericalangle$~\cite{Courtoy:2012ry,Radici:2018iag} so that we can now use it as a tool.

Using $H_1^\sphericalangle$ makes it possible to access collinear PDFs describing transversely polarized partons without introducing additional convolutions over the intrinsic transverse momentum in the initial and final state. The single-hadron equivalent in SIDIS, the Collins effect, mixes the transverse momentum dependence of the extracted PDF with the FF  and the PDF can only be extracted in a factorization framework that takes the intrinsic transverse momenta of the partons into account, the so-called TMD framework. Here TMD stands for Transverse Momentum Dependent. This is much less well understood as the collinear framework and scale evolution requires additional non-perturbative input \cite{Collins:2011zzd}. Furthermore, for a model independent extraction, the transverse momentum dependence of FF and PDF has to be known.
An example for the success of using DiFFs, is the extraction of the transversity PDF $h_1$ from a global fit, which was presented by M. Radici during this session~\cite{Radici:2018iag}. Unlike the extraction from single hadrons, this fit can incorporate $pp$ data~\cite{Adamczyk:2015hri}\cite{Abdelwahab:2014cvd} as well.

As mentioned above, having an additional degree of freedom allows the existence of DiFFs that have no correspondence in single-hadron fragmentation. An exciting example is the DiFF $G_1^\perp$, which describes the azimuthal dependence of an unpolarized hadron pair on the helicity of the outgoing quark~\cite{Boer:2003ya}. Similar to, {\textit e.g.,} the Boer-Mulders effect, this effect needs intrinsic transverse momentum acquired in the fragmentation process. In quark-jet models~\cite{Matevosyan:2017alv}, which describe in the transverse case the observed $H_1^\sphericalangle$ well, the intrinsic transverse momentum is acquired in the quark-to-quark splitting in the fragmentation process, and through the associated spin transfer the recoiling quarks acquire a non-zero transverse polarization. 

These polarizations are correlated, leading to an effect which is predicted to be significant, with the magnitude of $G_1^\perp$ about 30\% of the magnitude of $H_1^\sphericalangle$~\cite{Matevosyan:2017alv}.  Since we know that $H_1^\sphericalangle$ is large, therefore it is reasonable to expect that we should be able to observe effects sensitive to $G_1^\perp$. Extracting $G_1^\perp$ gives an exciting opportunity to study spin-momentum dynamics in hadronization. Fragmentation functions are not accessible on the lattice~\cite{Metz:2016swz}, therefore this channel presents an unique opportunity to extend our knowledge of hadronization.
The DiFF $G_1^\perp$ can be accessed in $A_\textrm{UL}$ as well as $A_\textrm{LU}$~\cite{Bacchetta:2002ux}, where $A_\textrm{UL}$ and $A_\textrm{LU}$ are longitudinal target- and beam-spin asymmetries defined in Sec.~\ref{sec:twist3}. In $A_\textrm{UL}$ it couples to the spin averaged PDF $f_1(x)$, whereas in $A_\textrm{UL}$ it couples to the helicity distribution $g_1(x)$, both of which have a significant magnitude. 
Some of the initial interest in $G_1^\perp$ was motivated by its connection to the so-called jet-handedness~\cite{Boer:2003ya}, which in turn might receive contributions from CP-violating QCD vacuum fluctuations~\cite{Efremov:2002qh}. This was one reason for the measurement at Belle~\cite{Abdesselam:2015nxn} which did not find a signal. However, after a revisit of the original calculation by another theory group~\cite{Matevosyan:2017liq,Matevosyan:2018icf}, it became clear this was due to a sign-mistake in the original calculation. A re-measurement of $G_1^\perp$ in $e^+e^-$ annihilation is planned to test the validity of factorization. Additionally, the comparison with the SIDIS measurement might be sensitive to parity violating vacuum fluctuations as discussed above.

\subsection{Accessing polarization dependence of quark-gluon interactions inside the nucleon through di-hadron correlations- the PDF $e(x)$}
\label{sec:twist3}
Recent experimental and theoretical progress has shown that understanding the gluonic degrees of freedom inside the nucleon is crucial to advance our knowledge of QCD. In fact, this is the main scientific motivation for the EIC~\cite{Accardi:2012qut}. 
Whereas the current physics program of the  EIC  focuses mainly on the role of the momentum (PDFs) and position distributions (GPDs), at JLab12 we can also access the dynamical role the gluons play inside the nucleon through twist-3 PDFs. That twist-3 PDFs are sensitive to quark-gluon interactions and that those should be large in any confined nucleon has been realized some time ago~\cite{Jaffe:1990qh}. The conventional definition of twist of the PDF is given by $d-s$, with $d$ being the mass dimension and $s$ the spin of the QCD operators in the PDF definition~\cite{Geyer:2000ma}. One often encounters a dynamical twist definition, given by the order $\nicefrac{1}{Q}$ the PDF contributes to the cross-section. Although, they do not coincide perfectly, both of these definitions are in common use and often interchangeably with twist-2 being the leading twist and twist-3 contributing at order $\nicefrac{1}{Q}$. 
Using the QCD equations of motion, the twist-3 PDFs of interest here can be decomposed into a $p_t$ weighted, TMD leading twist part, a current quark mass dependent part and a pure twist-3 part~\cite{Efremov:2002qh}, where the pure twist-3 part contains the interaction dependent part describing the coherent scattering of quark-gluon pairs with the quark that is probed in the hard scattering.
Even though twist-3 contributions are suppressed by one order of $\nicefrac{1}{Q}$,  this factor is not large at JLab12 and twist-3 PDFs themselves can be large. In fact, evidence of twist-3 PDFs has been seen in preliminary results from COMPASS and CLAS6~\cite{Sirtl:2017rhi,Pereira:2014hfa,Pisano:2014ila}. The effects are of the same order of magnitude than those of leading twist functions. This opens up the exciting possibility to measure long-range quark gluon correlations in the nucleon. At the same time it is also necessary to measure these effects since they are a significant background to any extraction of leading twist PDFs at JLab.
In the collinear case, there are only three non-vanishing twist-3 PDFs, $e(x)$, $g_T(x)$ and $h_L(x)$. Here, we are mainly interested in the PDF {\bf $e(x)$}. It is pure twist-3 and asymmetries connected to $e(x)$ have been shown to be significant in preliminary results from COMPASS and CLAS6~\cite{Sirtl:2017rhi,Pisano:2014ila}. The PDF $e(x)$ is connected to many fundamental QCD quantities~\cite{Efremov:2002qh}, but here we only want to highlight its interpretation in terms of gluons interacting with the struck quark, as discussed above. In this way, $e(x)$ can be interpreted as the transverse polarization dependence of the force that the struck quark in an unpolarized nucleon experiences right after the scattering~\cite{Burkardt:2008ps}. Of course this is extremely interesting, given the role that final-state-interactions play for our understanding of single spin asymmetries in SIDIS and the gauge structure of QCD~\cite{Burkardt:2003uw,Ji:2006ub}.
The PDF $e$ has been calculated in several models,{\textit e.g.} the MIT-bag model~\cite{Jaffe:1991ra} and the diquark spectator model~\cite{Jakob:1997wg}, which have been quite successful predicting other PDFs. 
Both models show a significant amplitude for both, $e$ and $h_L$. 
We will use calculations based on these models for the projections shown in Sec.~\ref{sec:plannedClas12Measurements}.

In di-hadron correlations $e(x)$ can be accessed in beam-spin asymmetries with an unpolarized target. 
The beam polarization $\lambda$ dependent part of the cross section reads~\cite{Bacchetta:2002ux}
\begin{equation}
\label{eq:LU}
d^7\sigma_{\textrm{LU}}=\frac{\alpha^2}{2\pi Q^2y}\lambda \sum e^2_a W(y) \sin\phi_R\frac{|\vec{R}_\perp|M}{QM_h}x e(x)H_1^\sphericalangle(z,M_h),
\end{equation}
where $e_a$ is the charge of quark flavor $a$, $y=\nicefrac{P^\mu q_\mu}{P^\mu l_\mu}$ and $W$ is a $y$ dependent kinematic factor. The subscript LU indicates that the beam is longitudinally polarized (L) and the target is unpolarized (U).
If we include the common factor $\sin\theta$ from the partial wave decomposition of the di-hadron FF, we can build the beam spin asymmetry $A_{LU}^{\sin\theta\sin\phi_R}=\frac{\int d \phi_R d\cos\theta \sin \phi_R(\sigma^+-\sigma^-}{d\phi_R d\cos\theta \sigma^++\sigma^-}$~\cite{Bacchetta:2003vn}, where $\sigma^{+/-}$ are the cross sections in~\eqref{eq:LU} with positive/negative beam helicity. The polar angle $\theta$ denotes the angle between $\vec{P}_h$ and the direction of emission of the two hadrons in their CMS system. In terms of the quantities of interest, the asymmetry is given by
\begin{equation}
\label{eq:ALU}
A_{LU}^{\sin\theta\sin\phi_R}(x,y,z,M_h,Q)=-\frac{W(y)}{A(y)}\frac{M}{Q}\frac{|\vec{R}_\perp|}{M_h}\frac{\sum e^2_a xe^q(x)H_{1,sp}^{\sphericalangle,a}(z,M_h)}{\sigma_a e^2_a f_1^a(x) D_{1,ss+pp}^q(z,M_h)}.
\end{equation}
Here, $A(y)$ is the kinematic factor associated with the unpolarized DiFF. 
Eqs.~\eqref{eq:LU} and ~\eqref{eq:ALU} omit the contribution from the twist-3 DiFF $\tilde{G}$, which is expected to vanish in the Wandzura-Wilczek approximation~\cite{Wandzura:1977qf}, but can also be eliminated in linear combination with $A_{LU}$~\cite{Bacchetta:2003vn}.

The simple expression~\eqref{eq:ALU} stands in contrast to the single-hadron asymmetry that is sensitive to $e(x)$. In fact, it is too long and complicated to be given here, and it mixes with other twist-3 PDFs, which would significantly complicate any extraction. Furthermore, in the single-hadron asymmetry, $e(x)$ couples to the TMD Collins function, and it is unclear if factorization is valid in the TMD picture at twist-3.

It is worth mentioning, that the target single spin asymmetry, $A_\textrm{UL}^{\sin\theta\sin\phi_R}$, which is defined analogues to  $A_\textrm{LU}^{\sin\theta\sin\phi_R}$, provides access to the twist-3 $h_L$, also coupled to $H_1^\sphericalangle$. Similarly to $e(x)$, the interaction dependent pure twist-3 part of this function is related to the gluon interaction with a transversely polarized struck quark, this time in a longitudinally polarized nucleon. However, unlike $e(x)$, in the decomposition of $h_L$ the twist-2 part does not vanish, and it reads
\begin{equation}
xh_L=\frac{m}{M}g_1(x) + 2 h_{1L}^{\perp(1)} + x \tilde{h}_L(x).
\end{equation}
Note that the factor $x$ is necessary due to a singularity at $x=0$. The first term of the {\textit r.h.s.} is the current quark mass ($m$) dependent part with the usual longitudinal PDF, the second term is the first term of the twist-2 TMD which is commonly known as "Worm-gear" and only the last term is the interaction dependent pure twist-3 part. Recently COMPASS~\cite{Sirtl:2017rhi} has shown a significant signal for $h_L$ in di-hadron asymmetries. Therefore, it would also be interesting to do a precision measurement at CLAS12.


\section{Transversely polarized $\Lambda$ production}
\label{sec:lambda}
Due to its self-analyzing weak decay, the inclusive production of polarized $\Lambda$ baryons has been of interest to the nuclear physics community for a long time. Historically, the observation of large transverse $\Lambda$ polarization in unpolarized $pp$ collisions~\cite{Bunce:1976yb} has been one of the key measurements that motivated a successful program to investigate transverse spin effects in nuclear physics~(see Refs.~\cite{Panagiotou:1989sv,Metz:2016swz} and references therein). Similar to DiFFs, being able to measure polarization of final state hadrons opens up the possibility to investigate spin-orbit correlations in fragmentation in a very similar framework to TMD PDFs~\cite{Pitonyak:2013dsu} and, on the other hand, access quark polarizations in the collinear framework if the spin transfer to the $\Lambda$ is known. Compared to DiFFs, $\Lambda$ baryons have a fixed angular momentum, so there is no proliferation of partial waves. Similar to the modified Gribov-Lipatov relation~\cite{Gribov:1972ri,Gribov:1972rt} between PDFs and FFs, it is reasonable to expect that the production of polarized $\Lambda$ baryons will provide new information about the structure of baryons. For example, the $p_t$ dependent production of transversely polarized $\Lambda^\uparrow$ from unpolarized quarks is the equivalent of the Sivers effect ($p_t$ dependence of the distribution of unpolarized quarks inside a transversely polarized nucleon). This is an example, how FFs can be classified analogously to polarized PDFs in the proton, once polarized spin-$\frac{1}{2}$ hadrons are measured in the final state. Beyond the inclusive production of $\Lambda^\uparrow$, there are many more interesting correlations to be explored in the production of polarized hadron pairs in $e^+e^-$ annihilation~\cite{Pitonyak:2013dsu}.
This is obviously very interesting given that proton and $\Lambda$ are related by a "simple" $u\leftrightarrow s$ interchange. In the twist-3 framework, the single spin asymmetry in $\Lambda$ leptoproduction can also provide information about the QCD mechanism involved in other single-spin asymmetries, notably in $pp$ collision, which are still not understood~\cite{Kanazawa:2015jxa}.
Lastly, polarized $\Lambda$ production is an important test of the gauge structure of QCD since the polarizing  FF $D_{1T}^{\perp}$ is a naive T-odd TMD like the Sivers PDF $f_{1}^\perp$. For these functions the Wilson link needed to construct a gauge invariant operator becomes more complex and can lead to a process dependence of the function~\cite{Collins:2002kn}, such as the sign change of the Sivers effect between SIDIS and Drell-Yan.
The exploration of this effect is a milestone (HP13) of the Nuclear Science Advisory Committee (NSAC) and the target of several experimental programs in Europe and the US~\cite{Adamczyk:2015gyk,Fazio:2016rtz,Aghasyan:2017jop}. In the fragmentation regime, it is thought that there is no such sign-change~\cite{Boer:2010ya}, however, so far this proposition has been difficult to test, since the only TMD T-odd FF that has been accessed in experiments, the Collins FF, is also chiral odd. Therefore it always appears in pairs with other chiral-odd objects, which makes the test for a sign change difficult because it may just cancel.  By measuring $D_{1T}^\perp$ for $\Lambda$ hyperons in SIDIS and $e^+e^-$ annihilation, the prediction about the gauge link structure of QCD in hadronization can be tested. This is deemed of similar importance as the aforementioned program on the Sivers effect~\cite{Boer:2010ya}.
\begin{figure}
\includegraphics[clip, trim=0cm 9.5cm 0cm 9.5cm,width=0.98\textwidth,height=0.25\textheight]{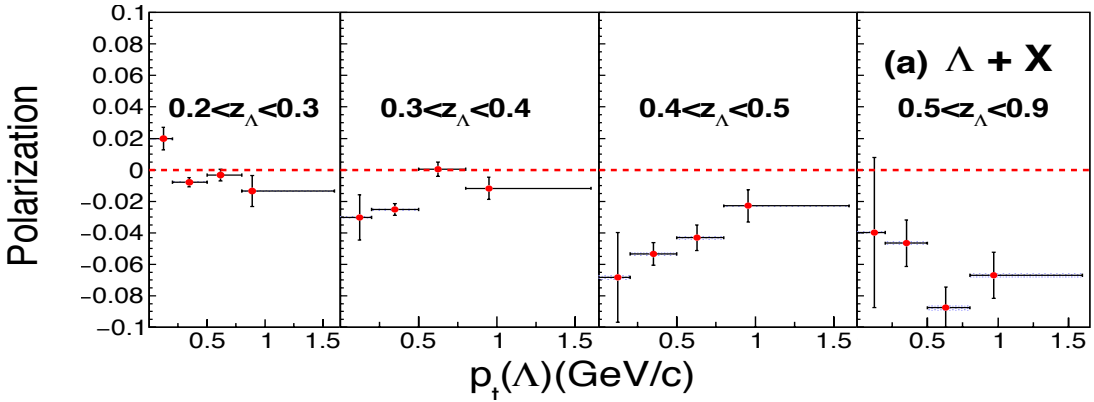}
\caption{Results on the transverse polarization of $\Lambda$ baryons in inclusive production in $e^+e^-$ annihilation. Each panel shows one bin in $z$ of the observed $\Lambda$ whereas the x-axis give the transverse momentum with respect to the thrust axis in the event. Figure taken from~\cite{Guan:2018ckx}\label{fig:belleLambda}}
\end{figure}

Recently, the Belle experiment showed the first measurement sensitive to $D_{1T}^{\perp\Lambda}$\cite{Guan:2018ckx}.
 Combining this measurement with a first observation in SIDIS would be an extremely important test of the gauge structure of QCD in hadronization. 

As often with first-of-a-kind measurements, the $e^+e^-$ results also pose new questions. In particular, the $p_T$ dependence shown in the mid-$z_\Lambda$ bins shown in Fig.~\ref{fig:belleLambda} is unexpected. It may be due to the flavor dependence of $D_{1T}^{\perp\Lambda}$ but a SIDIS measurement is needed for a more definitive answer.

\section{The CLAS12 Experiment at the upgraded {\mbox{CEBAF}} facility at JLAB}
\label{sec:clas12}
The newly upgraded  Continuous Electron Beam Accelerator Facility(CEBAF)~\cite{Leemann:2001dg} at JLab provides a  design luminosity of $10^{35}$cm$^{-2}$s$^{-1}$, corresponding to 0.1 pb$^{-1}$s$^{-1}$. Beam is delivered to four experimental halls. 
The CEBAF Large Acceptance Spectrometer (CLAS12)~\cite{CLAS12TDR} is located in experimental Hall B
Approved experiments at CLAS12 use a 11~GeV polarized electron beam that is scattered off an unpolarized or longitudinally polarized target. The target materials used in the unpolarized case are $H_2$ and $D_2$ and in the longitudinally polarized case $NH_3$ and $ND_3$, enabling a flavor separation of the measurements. There are also plans to take data with a transversely polarized $HD$ target ($HDIce$~\cite{Wei:2012ype}).

The CLAS12 detector has a wide acceptance with nearly $2\pi$ coverage in azimuth. It is separated into a forward detector for the detection of high momentum particles in the polar-angle range $5\degree-40\degree$, using a toroidal magnet with a peak field of 3.6T and a central detector part that covers a polar-angle of $35\degree-125\degree$ and which uses the magnetic field of a solenoid with a 5T field.
The kinematic coverage is $0.05 < x < 0.8 $ and $ 1.0 $ GeV$^2 < $Q$^2 < $10.5 GeV$^2$, where the lower bound is given by the requirement $W>2.0$ GeV to exclude the resonance region. The large kinematic coverage, combined with the integrated luminosity expected to be delivered, will allow us to study the $x$, $Q^2$ and $z$ dependence of our observables simultaneously which will be important for the physics analysis as well as for systematic studies. For the measurements discussed here we concentrate mainly on the forward detector for two reasons. First, this is were we expect to observe most of the hadrons produced in the current fragmentation region. Second, the forward detector has, due to its instrumentation and geometry, a very good angular resolution and PID. The later consists of a time-of-flight (TOF) detector as well as a low- and a high-threshold Cherenkov detector (LTCC, HTCC). The combined system is very efficient in identifying the scattered electron, however $\pi/K$ separation works reliably only to a momentum of about 3~GeV~\cite{CLAS12TDR}. Since in particular for SIDIS measurements the high-momentum region, corresponding to high $x$ and high $z$ is interesting, since this is where the largest effects are measured, an upgrade of the PID with a Ring Imaging Cherenkov Counter (RICH), that would replace the LTCC, is underway. At this point one of the six CLAS12 sectors is instrumented. 

\section{Planned Measurements at CLAS12}
\label{sec:plannedClas12Measurements}
\begin{figure}
\includegraphics[clip, trim=0cm 3.5cm 0cm 11.5cm,width=0.98\textwidth,height=0.25\textheight]{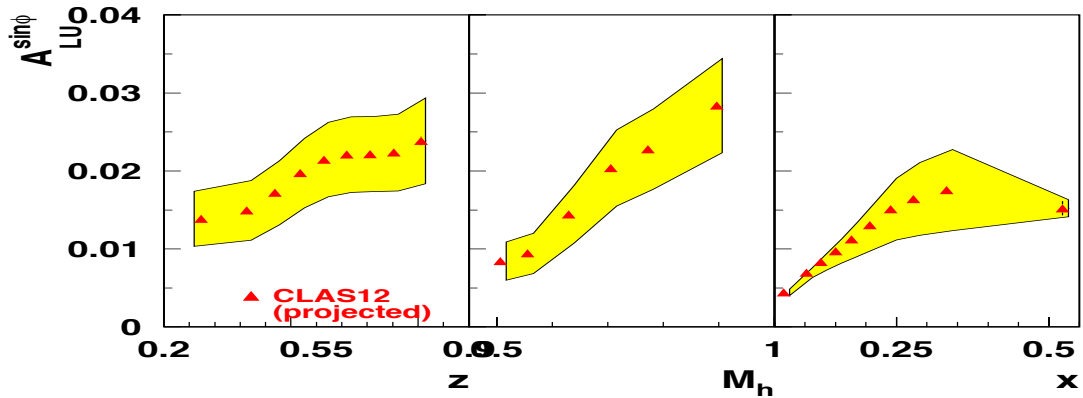}\\
\caption{Projections for $A_{LU}$ for $\pi^+\pi^-$ pairs using 30 days $NH_3$ data. The yellow bands shown in the figure correspond to the difference between the models used and do not reflect the systematic uncertainty of the measurement or the true theoretic uncertainty. See text for description. Figure taken from~\cite{proposal}\label{fig:diHadProjectionsProton}}
\end{figure}

The JLab PAC approved 60 days of running with an unpolarized liquid $H_2$ and liquid $D_2$ target (120 days total) are approved as well as 120 days with longitudinally polarized $NH_3$ and $ND_3$ targets. Using these datasets, an extraction of the beam and target spin asymmetries described in Eqs.~\eqref{eq:LU} and~\eqref{eq:ALU} will be extracted. Figure~\ref{fig:diHadProjectionsProton} shows the projections for $A_{LU}$ using about half the the dataset. The limits of the theoretical projections are given by the MIT bag model and di-quark spectator model discussed earlier, with the projected points plotted at the average value of these models. Even with this reduced dataset, the expected uncertainties are very small, allowing a precise extraction of the underlying twist-3 PDF $e$. In addition to the beam and target spin asymmetries for di-hadron pairs, a measurement of transversely polarized hyperons in the current fragmentation region is planned. Here we use a relatively weak definition of the current fragmentation region as particles having an $x$-Feynman larger than zero. The studies of this channel are still at an initial state, however it is expected that a first significant measurement of the transverse polarization of $\Lambda$ hyperons can be made.

\section{Planned Measurements at Belle II}
\label{sec:PlannedBelle}
In order to learn about the nucleon structure by extracting PDFs from cross-sections, FFs have to determined as well. 
From $e^+e^-$ annihilation data, FFs can be determined independently of PDFs, which proved to be crucial in particular for the extraction of quark-polarization dependent FFs. 
 Pioneering measurements sensitive to di-hadron and $\Lambda^\uparrow$ fragmentation functions  channels have been performed by the Belle experiment~\cite{Vossen:2011fk,Seidl:2015lla,Seidl:2017qhp,Guan:2018ckx}. Belle was a detector at the B-factory KEKB, located at the Japanese Institute for High Energy Physics (KEK), in Tsukuba. It was upgraded to Belle II, to take advantage of an increase of luminosity by a factor 40 from the upgrade of the accelerator to SuperKEKB with a target luminosity of $8\times 10^{35}/cm^2/s$. For more details on the experiment see {\textit e.g.}~\cite{Abe:2010gxa,b2tip}.
 Using Belle II data, it is planned to extract asymmetries sensitive to $G_1^\perp$ for pairs of pions and kaons following Ref.~\cite{Matevosyan:2018icf}. With the Belle II data, it will also be possible to revisit the production of $\Lambda^\uparrow$ and possibly even measure $\Lambda^\uparrow$-$\Lambda^\uparrow$ correlations.





\begin{thebibliography}{99}

\bibitem{Aidala:2012mv} 
  C.~A.~Aidala, S.~D.~Bass, D.~Hasch and G.~K.~Mallot,
  Rev.\ Mod.\ Phys.\  {\bf 85}, 655 (2013)
  doi:10.1103/RevModPhys.85.655
  [arXiv:1209.2803 [hep-ph]].
\bibitem{Sirtl:2017rhi} 
  S.~Sirtl,
  arXiv:1702.07317 [hep-ex].

  
  \bibitem{Vossen:2011fk} 
  A.~Vossen {\it et al.} [Belle Collaboration],
  Phys.\ Rev.\ Lett.\  {\bf 107}, 072004 (2011)
  doi:10.1103/PhysRevLett.107.072004
  [arXiv:1104.2425 [hep-ex]].

  
  \bibitem{Courtoy:2012ry} 
  A.~Courtoy, A.~Bacchetta, M.~Radici and A.~Bianconi,
  Phys.\ Rev.\ D {\bf 85}, 114023 (2012)
  doi:10.1103/PhysRevD.85.114023
  
  
  
  
  \bibitem{Radici:2018iag} 
  M.~Radici and A.~Bacchetta,
  Phys.\ Rev.\ Lett.\  {\bf 120}, no. 19, 192001 (2018)
  doi:10.1103/PhysRevLett.120.192001
  [arXiv:1802.05212 [hep-ph]].
  
\bibitem{Adamczyk:2015hri} 
  L.~Adamczyk {\it et al.} [STAR Collaboration],
  Phys.\ Rev.\ Lett.\  {\bf 115}, 242501 (2015)
  doi:10.1103/PhysRevLett.115.242501
  [arXiv:1504.00415 [hep-ex]].
  
  \bibitem{Abdelwahab:2014cvd} 
  L.~Adamczyk {\it et al.} [STAR Collaboration],
  Phys.\ Lett.\ B {\bf 751}, 233 (2015)
  doi:10.1016/j.physletb.2015.10.037
  [arXiv:1410.3524 [nucl-ex]].
\bibitem{Collins:2011zzd} 
  J.~Collins,
  Camb.\ Monogr.\ Part.\ Phys.\ Nucl.\ Phys.\ Cosmol.\  {\bf 32}, 1 (2011).
  
\bibitem{Boer:2003ya} 
  D.~Boer, R.~Jakob and M.~Radici,
  Phys.\ Rev.\ D {\bf 67}, 094003 (2003)
  Erratum: [Phys.\ Rev.\ D {\bf 98}, no. 3, 039902 (2018)]
  doi:10.1103/PhysRevD.98.039902, 10.1103/PhysRevD.67.094003
  [hep-ph/0302232].
  
\bibitem{Matevosyan:2017alv} 
  H.~H.~Matevosyan, A.~Kotzinian and A.~W.~Thomas,
  Phys.\ Rev.\ D {\bf 96}, no. 7, 074010 (2017)
  doi:10.1103/PhysRevD.96.074010
  [arXiv:1707.04999 [hep-ph]].
  
\bibitem{Metz:2016swz} 
  A.~Metz and A.~Vossen,
  Prog.\ Part.\ Nucl.\ Phys.\  {\bf 91}, 136 (2016)
  doi:10.1016/j.ppnp.2016.08.003
  [arXiv:1607.02521 [hep-ex]].
  
\bibitem{Bacchetta:2002ux} 
  A.~Bacchetta and M.~Radici,
  Phys.\ Rev.\ D {\bf 67}, 094002 (2003)
  doi:10.1103/PhysRevD.67.094002
  [hep-ph/0212300].
  
\bibitem{Efremov:2002qh} 
  A.~V.~Efremov and P.~Schweitzer,
  JHEP {\bf 0308}, 006 (2003)
  doi:10.1088/1126-6708/2003/08/006
  [hep-ph/0212044].
  
\bibitem{Abdesselam:2015nxn} 
  A.~Abdesselam {\it et al.} [Belle Collaboration],
  arXiv:1505.08020 [hep-ex].
    
  \bibitem{Matevosyan:2017liq} 
  H.~H.~Matevosyan, A.~Kotzinian and A.~W.~Thomas,
  Phys.\ Rev.\ Lett.\  {\bf 120}, no. 25, 252001 (2018)
  doi:10.1103/PhysRevLett.120.252001
  [arXiv:1712.06384 [hep-ph]].

\bibitem{Matevosyan:2018icf} 
  H.~H.~Matevosyan, A.~Bacchetta, D.~Boer, A.~Courtoy, A.~Kotzinian, M.~Radici and A.~W.~Thomas,
  Phys.\ Rev.\ D {\bf 97}, no. 7, 074019 (2018)
  doi:10.1103/PhysRevD.97.074019
  [arXiv:1802.01578 [hep-ph]].

  
\bibitem{Accardi:2012qut} 
  A.~Accardi {\it et al.},
  Eur.\ Phys.\ J.\ A {\bf 52}, no. 9, 268 (2016)
  doi:10.1140/epja/i2016-16268-9
  [arXiv:1212.1701 [nucl-ex]].
  
  
\bibitem{Jaffe:1990qh} 
  R.~L.~Jaffe and X.~D.~Ji,
  Phys.\ Rev.\ D {\bf 43}, 724 (1991).
  doi:10.1103/PhysRevD.43.724
  
\bibitem{Geyer:2000ma} 
  B.~Geyer and M.~Lazar,
  Phys.\ Rev.\ D {\bf 63}, 094003 (2001)
  doi:10.1103/PhysRevD.63.094003
  [hep-ph/0009309].

\bibitem{Pereira:2014hfa} 
  S.~A.~Pereira [CLAS Collaboration],
  PoS DIS {\bf 2014}, 231 (2014).
  doi:10.22323/1.203.0231
\bibitem{Pisano:2014ila} 
  S.~Pisano [CLAS Collaboration],
  EPJ Web Conf.\  {\bf 73}, 02008 (2014).
  doi:10.1051/epjconf/20147302008

\bibitem{Burkardt:2008ps} 
  M.~Burkardt,
  Phys.\ Rev.\ D {\bf 88}, 114502 (2013)
  doi:10.1103/PhysRevD.88.114502
  [arXiv:0810.3589 [hep-ph]].
  
\bibitem{Burkardt:2003uw} 
  M.~Burkardt,
  Nucl.\ Phys.\ A {\bf 735}, 185 (2004)
  doi:10.1016/j.nuclphysa.2004.02.008
  [hep-ph/0302144].
  
\bibitem{Ji:2006ub} 
  X.~Ji, J.~W.~Qiu, W.~Vogelsang and F.~Yuan,
  Phys.\ Rev.\ Lett.\  {\bf 97}, 082002 (2006)
  doi:10.1103/PhysRevLett.97.082002
  [hep-ph/0602239].
  
\bibitem{Jaffe:1991ra} 
  R.~L.~Jaffe and X.~D.~Ji,
  Nucl.\ Phys.\ B {\bf 375}, 527 (1992).
  doi:10.1016/0550-3213(92)90110-W
  
\bibitem{Jakob:1997wg} 
  R.~Jakob, P.~J.~Mulders and J.~Rodrigues,
  Nucl.\ Phys.\ A {\bf 626}, 937 (1997)
  doi:10.1016/S0375-9474(97)00588-5
  [hep-ph/9704335].
  
\bibitem{Bacchetta:2003vn} 
  A.~Bacchetta and M.~Radici,
  Phys.\ Rev.\ D {\bf 69}, 074026 (2004)
  doi:10.1103/PhysRevD.69.074026
  [hep-ph/0311173].
  
\bibitem{Wandzura:1977qf} 
  S.~Wandzura and F.~Wilczek,
  Phys.\ Lett.\  {\bf 72B}, 195 (1977).
  doi:10.1016/0370-2693(77)90700-6
  
\bibitem{Bunce:1976yb} 
  G.~Bunce {\it et al.},
  Phys.\ Rev.\ Lett.\  {\bf 36}, 1113 (1976).
  doi:10.1103/PhysRevLett.36.1113
  
\bibitem{Panagiotou:1989sv} 
  A.~D.~Panagiotou,
  Int.\ J.\ Mod.\ Phys.\ A {\bf 5}, 1197 (1990).
  doi:10.1142/S0217751X90000568
  
\bibitem{Pitonyak:2013dsu} 
  D.~Pitonyak, M.~Schlegel and A.~Metz,
  Phys.\ Rev.\ D {\bf 89}, no. 5, 054032 (2014)
  doi:10.1103/PhysRevD.89.054032
  [arXiv:1310.6240 [hep-ph]].
  
  
  
\bibitem{Gribov:1972ri} 
  V.~N.~Gribov and L.~N.~Lipatov,
  Sov.\ J.\ Nucl.\ Phys.\  {\bf 15}, 438 (1972)
  [Yad.\ Fiz.\  {\bf 15}, 781 (1972)].

\bibitem{Gribov:1972rt} 
  V.~N.~Gribov and L.~N.~Lipatov,
  Sov.\ J.\ Nucl.\ Phys.\  {\bf 15}, 675 (1972)
  [Yad.\ Fiz.\  {\bf 15}, 1218 (1972)].
  
\bibitem{Kanazawa:2015jxa} 
  K.~Kanazawa, A.~Metz, D.~Pitonyak and M.~Schlegel,
  Phys.\ Lett.\ B {\bf 744}, 385 (2015)
  doi:10.1016/j.physletb.2015.04.011
  [arXiv:1503.02003 [hep-ph]].
  
\bibitem{Collins:2002kn} 
  J.~C.~Collins,
  Phys.\ Lett.\ B {\bf 536}, 43 (2002)
  doi:10.1016/S0370-2693(02)01819-1
  [hep-ph/0204004].


\bibitem{Adamczyk:2015gyk} 
  L.~Adamczyk {\it et al.} [STAR Collaboration],
  Phys.\ Rev.\ Lett.\  {\bf 116}, no. 13, 132301 (2016)
  doi:10.1103/PhysRevLett.116.132301
  [arXiv:1511.06003 [nucl-ex]].
  
\bibitem{Fazio:2016rtz} 
  S.~Fazio [STAR Collaboration],
  PoS DIS {\bf 2016}, 213 (2016)
  doi:10.22323/1.265.0213
  [arXiv:1607.01676 [nucl-ex]].
  
\bibitem{Aghasyan:2017jop} 
  M.~Aghasyan {\it et al.} [COMPASS Collaboration],
  Phys.\ Rev.\ Lett.\  {\bf 119}, no. 11, 112002 (2017)
  doi:10.1103/PhysRevLett.119.112002
  [arXiv:1704.00488 [hep-ex]].
  
\bibitem{Boer:2010ya} 
  D.~Boer, Z.~B.~Kang, W.~Vogelsang and F.~Yuan,
  Phys.\ Rev.\ Lett.\  {\bf 105}, 202001 (2010)
  doi:10.1103/PhysRevLett.105.202001
  [arXiv:1008.3543 [hep-ph]].
  
\bibitem{Leemann:2001dg} 
  C.~W.~Leemann, D.~R.~Douglas and G.~A.~Krafft,
  Ann.\ Rev.\ Nucl.\ Part.\ Sci.\  {\bf 51}, 413 (2001).
  doi:10.1146/annurev.nucl.51.101701.132327
  
\bibitem{Wei:2012ype} 
  X.~Wei {\it et al.},
  J.\ Phys.\ Conf.\ Ser.\  {\bf 400}, no. 5, 052042 (2012).
  doi:10.1088/1742-6596/400/5/052042
  
  
  
  
  
    \bibitem{Guan:2018ckx} 
  Y.~Guan {\it et al.} [Belle Collaboration],
  arXiv:1808.05000 [hep-ex].
\bibitem{proposal}
JLab proposal PR12-11-109, available at \url{https://www.jlab.org/exp_prog/proposals/11/PR12-11-109.pdf}


  
\bibitem{Seidl:2015lla} 
  R.~Seidl {\it et al.} [Belle Collaboration],
  Phys.\ Rev.\ D {\bf 92}, no. 9, 092007 (2015)
  doi:10.1103/PhysRevD.92.092007
  [arXiv:1509.00563 [hep-ex]].
  
\bibitem{Seidl:2017qhp} 
  R.~Seidl {\it et al.} [Belle Collaboration],
  Phys.\ Rev.\ D {\bf 96}, no. 3, 032005 (2017)
  doi:10.1103/PhysRevD.96.032005
  [arXiv:1706.08348 [hep-ex]].

\bibitem{CLAS12TDR}
CLAS12 TDR available at \url{https://www.jlab.org/Hall-B/clas12_tdr.pdf}.
\bibitem{Abe:2010gxa} 
  T.~Abe {\it et al.} [Belle-II Collaboration],
  arXiv:1011.0352 [physics.ins-det].
\bibitem{b2tip}
KEK Preprint 2018-27,BELLE2-PUB-PH-2018-001, FERMILAB-PUB-18-398-T, JLAB-THY-18-2780, INT-PUB-18-047, UWThPh 2018-26

\end{thebibliography}
\end{document}